\documentclass[aoas, nameyear,dvips]{arximspdf}
\usepackage{graphicx}

\doi{10.1214/10-AOAS347}
\volume{4}
\issue{4}
\pubyear{2010}
\firstpage{2114}
\lastpage{2125}

\makeatletter
\newtheorem{theorem}{Theorem}[section]
\newtheorem{lemma}{Lemma}
\makeatother

\begin{document}
\begin{frontmatter}

\title{Detection of treatment effects by covariate-adjusted expected shortfall\thanksref{TX1}}
\thankstext{TX1}{Supported in part by the National Science
Foundation (USA) Award DMS-06-04229, National Institute of Health (USA)
Grant R01GM080503-01A1, NNSF of China Grant 10828102, and a Changjiang Visiting Professorship at the Northeast Normal
University, China.}
\runtitle{Detection of treatment effects}

\begin{aug}
\author[A]{\fnms{Xuming} \snm{He}\ead[label=e1]{x-he@illinois.edu}\corref{}},
\author[A]{\fnms{Ya-Hui} \snm{Hsu}\ead[label=e2]{yhsu6@illinois.edu}}
\and
\author[B]{\fnms{Mingxiu} \snm{Hu}\ead[label=e3]{mingxiu.hu@mpi.com}}
\runauthor{X. He, Y.-H. Hsu and M. Hu}
\affiliation{University of Illinois at Urbana-Champaign, University of Illinois at Urbana-Champaign and Millennium Pharmaceuticals, Inc.}
\address[A]{
 X. He\\
 Y.-H. Hsu\\
University of Illinois at Urbana-Champaign\\
725 South Wright Street, Champaign\\
Illinois 61820\\
USA\\\printead{e1}\\
\hphantom{E-mail: }\printead*{e2}} %adresu isvedimo komanda gale!
\address[B]{
M. Hu\\
Millennium Pharmaceuticals, Inc.\\
35 Landsdowne Street, Cambridge\\
Massachusetts 02139\\
USA\\\printead{e3}}
\end{aug}

% HISTORY:
\received{\smonth{10} \syear{2009}}
\revised{\smonth{3} \syear{2010}}

% ABSTRACT
\begin{abstract}
The statistical tests that are commonly used for detecting mean or median treatment
effects suffer from low power when the two distribution functions
differ only in the upper (or lower) tail, as in the assessment of
the Total Sharp Score (TSS) under different treatments for
rheumatoid arthritis. In this article, we propose a more powerful
test that detects treatment effects through the expected shortfalls.
We show how the expected shortfall can be adjusted for covariates,
and demonstrate that the proposed test can achieve a substantial sample size reduction over the conventional tests on the mean effects.
\end{abstract}

% KEYWORDS

\end{frontmatter}

%s1 ###
\section{Introduction}
We consider the problem of testing the hypothesis of
no treatment effect against a class of alternatives where the two
outcome distributions differ only or mainly in the right tail. As
demonstrated in some recent trials of rheumatoid arthritis therapies
in \citet{vanderHeijde2006} and \citet{Kremer2006}, the
changes in Total Sharp Scores, the primary measurements of the
treatment effects on prevention of structural damage, are nearly identical for most therapies for
nearly 75\% of the patient population, but the difference lies in the most
challenging 25\% of the patient population where a less effective treatment loses
its efficacy, resulting in a heavy right tail in its outcome distribution.
The two-sample $t$-test or its regression counterpart in
covariate-adjusted linear models is commonly used for detecting the
treatment effects, but due to skewness and heavy-tails of the
distributions, the test does not have satisfactory power.
Nonparametric tests on the median differences, for example, would
fare even worse in such cases, because the median differences are
often negligible among those therapies.

A natural test in this type of applications is to
focus on the average in one tail, or the expected tail loss (aka
expected shortfall). In finance, this is often referred to as the conditional
value at risk (CVaR), for measuring the risk of a portfolio. In our context, a
treatment is said to be more effective if it has a smaller expected
shortfall, where the expected shortfall is defined to be the
conditional mean of the outcome (e.g., change in Total Sharp Score)
above the $\tau$th quantile. In this paper, $\tau$ will be taken to be
a user-specified value (e.g., 0.75), but a good choice of $\tau$ clearly depends on the area of
applications. In finance, the most relevant choices of $\tau$ fall
above 0.90.

A two-sample comparison of the expected shortfalls is not difficult,
as it falls into the well-known theory of the $L$-statistic. In fact, there are also
a large number of other tests that one can use to compare tails of
two outcome distributions, but few have been developed to adjust for
covariates. The purpose of this paper is to
develop a simple test for testing the hypothesis on the
treatment effect adjusting for certain covariates; the proposed test uses the
\textit{COV}ariate-adjusted \textit{E}xpected \textit{S}hortfall (COVES).

Our work starts with a brief
introduction to our motivating example on the TSS for rheumatoid
arthritis therapies in Section \ref{sec2}. In Section \ref{sec3}, we  propose an
appropriate treatment effect size of covariate-adjusted expected
shortfall, followed by a new test for detecting differences in the treatment
effects. The large sample theory
for the proposed test is given here. In Section \ref{sec4}, we compare the proposed COVES test with the
$t$-test based on the least squares regression in empirical power. In
particular, we show that when the outcome distributions resemble
those of the TSS, the COVES test has a clear advantage in
reducing sample sizes in clinical trials.  The basic
idea and methodology developed in this paper apply to other
problems of comparing two covariate-adjusted tails of outcome distributions.
In Section \ref{sec5}, we provide a diagnostic tool that can be used to gauge the
need for the proposed test and to guide the selection of $\tau$. Section \ref{sec6} concludes
the paper with some additional remarks about the COVES test.

%s2 ###
\section{A primer on total sharp scores}\label{sec2}

Rheumatoid arthritis (RA) is a chronic disabling disease that causes
destruction of joint cartilage and erosion of adjacent bones. In RA
clinical trials, TSS is used to measure the treatment effect of RA
drugs on prevention of structural damage to the joints. It consists
of two components, erosion score and score for joint space narrowing
(JSN), which are obtained through examination of hand and/or feet
joints with radiographic methods. The first description of TSS is
given by \citet{Sharp1971}, but TSS has been modified in later
studies. The example presented in this paper is based on van der
Heijde's modification of TSS scoring system [\citet{vanderHeijde2000}],
which is based on examination of 16 areas for erosions and 15 for
joint space narrowing in each hand. The erosion score per joint
ranges from 0 to 5 with 0 representing a normal condition and 5 the
most severe disease, and thus the total erosion score ranges from 0
to 160 (16 areas by 2 hands by 5). The JSN score ranges from 0 to 4
per joint with higher score representing more severe disease, which
leads to a range of 0 to 120 (15 areas by 2 hands by 4) for the
total JSN score. Therefore, the range of TSS is 0 -- 280. The
primary interest is the change from baseline in TSS in one or two
years.

\begin{figure}

\includegraphics{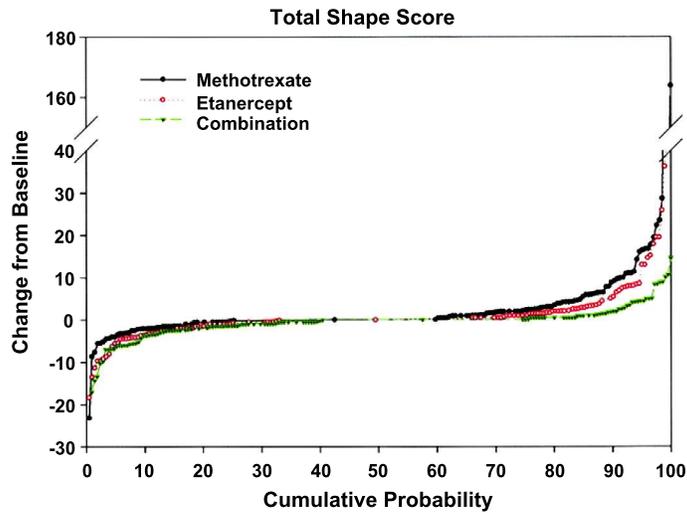}

\caption{This figure, reproduced from van der Heijde et al.
(\protect\citeyear{vanderHeijde2006}), shows that the changes in TSS in the TEMPO trials
differ mostly in the upper tails.}\label{fig1}
\end{figure}

The change in TSS has a highly skewed distribution under any known
treatment. In the TEMPO trial [\citet{vanderHeijde2006}] comparing
Methotrexate, Etanercept, and the combination therapy of Etanercept
and Methotrexate, the three treatments are similarly effective for
about 75\% of the patients whose conditions improved or showed no or
little progression from the baseline; see Figure~\ref{fig1}.
%%
%f1 ###
Medians for all three groups are around 0. Treatment differences come from the 25\% of the patients with the most progressive diseases. In other
words, the differences in treatment effects are not attributed to a
location-scale change in the distributions. The distributions of
clinical data from several other major RA trials [\citet{Kremer2006}; \citet{Keystone2004}; \citet{Lipsky2000}] showed similar characteristics.

It is clear that the distributions for the changes in TSS are far
from normal, and the $t$-test is expected to lose power due to skewness and
heavier-tails that are evident in the data. Nonparametric tests on
the median differences would fare even worse, because the median
differences of those treatments are essentially nonexistent. Researchers in some
trials have considered the chi-square tests on the proportion of
patients with little disease progression by dichotomizing TSS, but
there has been no agreeable cutoff point for dichotomization. In
fact, the power of the chi-square test depends rather critically on
the cutoff point. In addition, it is difficult to perform the
chi-square test when a covariate needs to be adjusted for. A~natural
quantity for distinguishing treatment effects is the expected
shortfall, which averages the changes in TSS in the upper tail. We
propose to use the regression quantile approach of \citet{KoenkerandBassett1978} to estimate the covariate-adjusted expected
shortfall.

Later in this paper, we use a recent observational study conducted
at Brigham and Women's Hospital and sponsored by Millennium
Pharmaceuticals Inc. and Biogen Idec as a basis for assessing the
performance of the proposed test. We take 150 subjects in the study,
who are under active treatment, and simulate a control group whose
outcome distribution is chosen to mimic the treatment difference
reported in other trials. For example, in the Adalimumab trial
[\citet{Keystone2004}], the variance of the treatment group (using
the drug Adalimumab 20~mg$/$kg) is about half of that in the control
group (using the drug Methotrexate) with a mean difference of $-1.9$.
In the Abatacept trial [\citet{Kremer2006}], the variance in the
Abatacept group is about one third of that in the control group. In
our simulation studies, we use the ratio of variances between 2:1
and 3:1 between two treatment groups.

%s3 ###
\section{Proposed test: COVES}\label{sec3}

We use a dummy variable $D$ as treatment indicator, $C$ as the
covariate of interest, and $Z$ as the outcome measure. For
simplicity of notation, we consider $C \in R$ as a univariate
covariate and $D$ taking values 0 or 1, but the work generalizes
readily for multivariate covariates and multiple treatments. As appropriate with
randomized trials, we assume that $C$ and $D$ are independent. We
model the $\tau$th quantile of $Z$ given $(D, C)$ as
%e1 ###
\begin{equation} \label{quant}
Q_Z(\tau|D,C)=\alpha(\tau) +\delta(\tau)D+\gamma(\tau)C ,
\end{equation}
where the coefficients $\alpha, \delta$, and $\gamma$ are
$\tau$-specific. In this paper, we use $\tau=0.75$ for empirical studies, but refer to
Section \ref{sec5} for guidance on the selection of $\tau$. We also refer to
\citet{Koenker2005} for details on the linear regression quantile
specification.

Given data $(Z_i, D_i, C_i )$ with $D_i =1$ for $i=1, \dots , m$
and $D_i=0$ for $i=m+1, \dots , m+n$, we can use the \textit{quantreg}
package in R to obtain the regression quantile coefficient $\hat
\alpha$, $\hat \delta$, and $\hat \gamma$. Then, let ${\hat e}_i =Z_i
-\hat\alpha  -\hat\delta D_i-\hat\gamma C_i$ as the residuals from
the $\tau$th regression quantile. By contrast, we also write ${
e}_i =Z_i -\alpha (\tau)  - \delta (\tau)  D_i- \gamma (\tau) C_i$,
which has zero as the $\tau$th conditional quantile given ($D_i,
C_i$) due to (\ref{quant}).

Let $Y_i = Z_i - \hat \gamma C_i$ be the \textit {covariate-adjusted outcome},
and define the empirical covariate-adjusted expected shortfall for
the two groups as
\[
\operatorname{COVES}_\tau (d )=\sum_{D_{i}=d}  w_{d, i}Y_i,\qquad d =0,1,
\]
where $w_{d,i}  = S_d^{-1} I( {\hat e}_i >0)$ and $S_d= \sum_{D_i=d}
I( {\hat e}_i >0)$. The quantity $\operatorname{COVES}_\tau (d )$ is the
average of the outcomes for group $d$ that are
above the $\tau$th covariate-adjusted quantile.

The proposed $\operatorname{COVES}$ test statistic for the hypothesis of no
difference between the two treatment groups is given as
%e2 ###
\begin{equation} \label{t1}
T_\tau (m,n)= \operatorname{COVES}_\tau (1 ) - \operatorname{COVES}_\tau (0 )  .
\end{equation}

Let ${\bar C}_\tau (d)$ and ${\bar e}_\tau (d)$ be the average of $C_i$
and $e_i$ in group $d$ that are above the $\tau$th regression quantile, that is,
\begin{eqnarray*}
{\bar C}_\tau (d)&=&S_d^{-1} \sum_{D_{i}=d} C_i I( {\hat e}_i >0 ),\\
{\bar e}_\tau (d)&=& S_d^{-1} \sum_{D_{i}=d} \bigl(Z_i-\alpha(\tau )-\delta
(\tau)D_i -\gamma(\tau )C_i \bigr)I( {\hat e}_i >0 ).
\end{eqnarray*}
Then, the test statistic (\ref{t1}) can be written as
%e3 ###
\begin{equation} \label{t2}
T_\tau(m,n) = \delta(\tau)-\bigl(\hat\gamma - \gamma(\tau )\bigr)\bigl({\bar C}_\tau(1)-{\bar C}_\tau(0)\bigr)
+\bigl({\bar e}_\tau(1)-{\bar e}_\tau(0)\bigr),
\end{equation}
which makes it relatively easy to establish the asymptotic normality
of the test statistic as $m, n \to \infty$.

To estimate the variance of $T_\tau(m,n)$, let $N_d=\sum_i I(D_i
=d)$, $f_i$ be the conditional density function of $e_i$ given
$(D_i, C_i)$ evaluated at 0, and
\[
C_i^* = C_i - N_d^{-1} \sum_i {C_i} I(D_i=d),
\]
as the orthogonal components $C$ relative to the treatment groups.
In more general problems, we can obtain $C^*$ by the Gram--Schmidt
orthogonalization of the design matrix. Furthermore, let
\begin{eqnarray*}
V_d &=& \sum_{D_i =d} \{ {\hat e}_i^2   I ( {\hat e}_i >0 ) \} -
N_d^{-1}\biggl[ \sum_{D_i =d} \{ {\hat e}_i   I ( {\hat e}_i >0 ) \} \biggr]^2 ,\\
\qquad U_f &=& \sum_i (f_i {C_i^*}^2),
\end{eqnarray*}
and
\begin{eqnarray} \label{var}
&&s_{m,n}^2 = (1- \tau)^{-2} (V_1/m^2 + V_0/n^2)\nonumber\\[-8pt]\\[-8pt]
&&\hphantom{s_{m,n}^2 =}{}+ \tau (1-\tau) \bigl({\bar C}_\tau(1)-{\bar C}_\tau(0)\bigr)^2 U_f^{-2} \biggl(\sum_i
{C_i^*}^2\biggr).\nonumber
\end{eqnarray}

\begin{theorem} \label{theorem}
Suppose that $\lim_{m,n\rightarrow\infty}(m+n)^{-1}U_f$ exists, $E|C_i|^3<\infty$, and $f_i$ are uniformly bounded
away from 0 and infinity. Under the null hypothesis that $F_{Z|C, D=1}=F_{Z|C, D=0}$, we have
\[
T_\tau^{\mathrm{COVES}} (m,n) /s_{m,n} \to N(0,1)\qquad\mbox{as}\,m,n \to
\infty.
\]
\end{theorem}

The proof of Theorem \ref{theorem} is given in the \hyperref[append]{Appendix}, but to use the
asymptotic normality for testing the null hypothesis of no treatment
effects, we need a consistent estimate of $U_f$. If $e_i$ in each
group (corresponding to $D_i=0$ or 1) follows a common distribution,
then a kernel density estimate can be used to estimate the common
density at 0 from ${\hat e}_i$ in the $d$th group. If the
conditional densities vary with $C_i$, it is not possible to
estimate each $f_i$ consistently, but $U_f$, a linear combination of
the $f_i$'s, can still be consistently estimated; see \citet{He2002} and \citet{Koenker2005} for more details. For the empirical
investigations in this paper, the proposed test is carried out using
a kernel density estimate, \textit{density}, in R on each treatment group.

%s4 ###
\section{Empirical investigations}\label{sec4}
In this section, we report some empirical power studies of the
proposed test based on Monte Carlo simulations. The first study is
constructed based on the data we obtained from a recent study on an
undisclosed therapy to treat RA at the Brigham and Women's Hospital
in Boston. The other studies are constructed with other types of
distributions in mind. Together, we find that the proposed
$\operatorname{COVES}$
test greatly outperforms the usual regression tests on the mean
differences when the group differences occur at one tail of the
distributions.

%s4.1 ###
\subsection{Targeted study on TSS}\label{4.1}
We use the empirical distributions, $F$, of the TSS changes of 150 patients in the Brigham and Women's Hospital study as the underlying distribution for the group $d=1$.
We take the baseline TSS as the covariate in the analysis, whose empirical distribution
for the group $d=1$
will be denoted as $G$.

\begin{figure}

\includegraphics{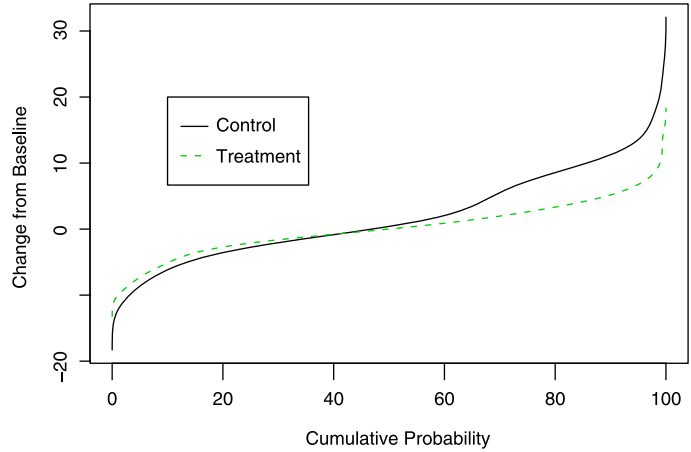}

\caption{Quantile function of the TSS
change shows that the groups differ mostly in the upper tails.}\label{fig2}
\end{figure}

The data from the control
group (with $d=0$) will be simulated as
\[
C= G^{-1} (u),\qquad Z = F^{-1}(u) + 8 |u-0.65|^{1/4} I(u>0.65),
\]
where $u$ is a uniform random number in (0, 1). Clearly, the control
group has a heavier right tail in its outcome, but the covariate $C$ has
the same distribution in both groups. In this setting, the variance of the control
group is about twice that of the treatment group.
Table \ref{tab1} and Figure \ref{fig2} summarize the
differences of the two groups.
%%%
%t1 ###
\begin{table}[b]
\caption{Differences in the $\tau$th quantiles and in the mean,
with the last column as the ratio of the variances between the
control group ($d=0$) and the treatment group ($d=1$)}\label{tab1}
\begin{tabular*}{\textwidth}{@{\extracolsep{\fill}}lccccccccc@{}}
\hline
$\bolds\tau$ &  \textbf{0.5} & \textbf{0.6} & \textbf{0.7}&\textbf{0.75}&\textbf{0.8}&\textbf{0.9}&\textbf{0.99}& \textbf{Mean} &\textbf{Variance ratio}\\
\hline
&0 &  0&  3.72& 4.53& 4.96& 5.64&6.02&1.74&2.03\\
\hline
\end{tabular*}
\end{table}
%%%
%f2 ###
%%%
%f3 ###
\begin{figure}

\includegraphics{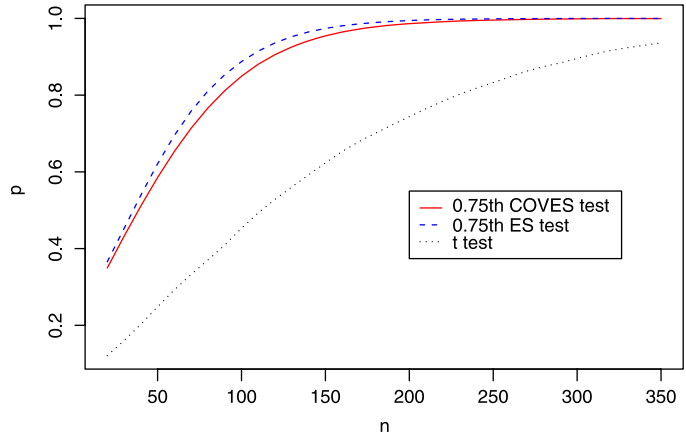}

\caption{Statistical powers of three tests in the
targeted study on TSS as functions of sample size $m=n$. The ES test
ignores the covariate in the model.}\label{fig3}
\end{figure}

The
power functions for the $\operatorname{COVES}$ test with $\tau=0.75$ and the $t$-test
from linear regression are shown in Figure \ref{fig3} with sample
sizes up to $m=n=350$. For comparison, we also include in the figure
the power curve for the test based on expected shortfalls (ES)
without adjusting for the baseline TSS.  Table \ref{tab2} provides
the sample sizes needed to reach a power of 0.90 in clinical trials
with $m=n$ as well as $m=2n$. It is common in clinical experiments
to allocate twice as many patients to the treatment group when the
treatment is believed to be effective. In this case, the baseline
TSS does not play a significant role, so the statistical power for
detecting the treatment effect has no gain by adjusting the
covariate in the analysis. However, the results show that the COVES
test is clearly outperforming the $t$-test, and the latter would
require a trial that is more than double in size.
%%
%t2 ###
\begin{table}\tablewidth=210 pt
\caption{Sample sizes needed to reach power 0.9. The cases of $m=n$
and $m=2n$ are included}\label{tab2}
\begin{tabular*}{210 pt}{@{\extracolsep{\fill}}lc@{}}
\hline
&\textbf{Sample size $\bolds{(m,n)}$}\\
\hline
$\operatorname{COVES}$ test ($\tau=0.75$) & (120, 120) or (172, 86) \\
$t$-test& (306, 306) or (450, 225) \\
\hline
\end{tabular*}
\end{table}

%s4.2 ###
\subsection{More simulation studies}\label{4.2}
We consider data generated from
%e4 ###
\begin{equation}\label{normalmodel}
Z_i=5+\gamma C_i+\{1 + \eta I(e_i >0) I(D_i=0)\} e_i,
\end{equation}
where $e_i\sim N(0,1)$, and $\eta$ is either 0 (under the null
hypothesis) or 1.35 (under the alternative hypothesis). The
coefficient $\gamma$ and the distribution for the covariate $C_i$
will be specified later. Clearly, the control group ($d=0$) has a
heavier right tail. When $\eta =1.35$, the error variance of the
control group ($d=0$) is about triple that of the treatment group
($d=1$) under this model. Table \ref{tab3} summarizes the
differences of the two groups under the alternative hypothesis.

%t3 ###
\begin{table}\tablewidth=270 pt
\caption{Difference of the two groups at $\eta=1.35$, with the last
column for the ratio of error variances}\label{tab3}
\begin{tabular*}{270 pt}{@{\extracolsep{\fill}}lcccccccc@{}}
\hline
$\bolds\tau$ & \textbf{0.5} & \textbf{0.6}& \textbf{0.7}&\textbf{0.75}&\textbf{0.8}&\textbf{0.9}& \textbf{Mean} &\textbf{Var ratio}\\
\hline
&0 & 0.34 & 0.70&0.91&1.13&1.72&0.54&\textbf{2.97}\\
\hline
\end{tabular*}
\end{table}

We will consider four scenarios for the effects of the covariate in
the analysis:
\begin{itemize}

\item \textit{Scenario 1, no covariate effect:} we take $C_i$ from
$N(2.5, 0.5^2)$, with $\gamma=0$.

\item \textit{Scenario 2, a common covariate effect:}  we take $C_i$
from $N(2.5, 0.5^2)$, with $\gamma=1$.

\item \textit{Scenario 3, a covariate distribution that varies with
treatment groups:}   we take $C_i$ from $N(2.5, 0.5^2)$ for $d=0$,
but from $N(3.0, 0.5^2)$ for $d=1$, with $\gamma=1$.

\item \textit{Scenario 4, a covariate distribution that has a scale
change across treatment groups:} we take $C_i$ from
$N(2.5, 0.5^2)$ for $d=0$, but from $N(2.5, 1.0)$ for $d=1$, with
$\gamma=1$.

\end{itemize}

Scenarios 3 and 4 are unlikely for randomized trials, but we include them in the
study to examine the robustness of the $\operatorname{COVES}$ test when the covariate distributions vary
to some extent with the treatment groups. The type I errors of the $\operatorname{COVES}$ test and the $t$-test under these
scenarios are controlled to stay close to the nominal level of 0.05. The
following table reports the type I errors at the sample size of
$m=n=50$. It also reports the sample sizes needed to reach power of
0.90 in each scenario under two design conditions: $m=n$ and $m=2n$,
respectively.
%t4 ###
\begin{table}
\caption{Simulation comparisons for the $\operatorname{COVES}$ test versus t-test
for linear models. The
sample sizes under two conditions $m=n$ and $m=2n$ are given}\label{tab4}
\begin{tabular*}{\textwidth}{@{\extracolsep{\fill}}lcccc@{}}
\hline
 & \multicolumn{2}{c}{$\bolds{\operatorname{COVES}}$ \textbf{test}} &
 \multicolumn{2}{c@{}}{\textbf{\textit{t}-test}}\\[ -7 pt]
 & \multicolumn{2}{c}{\hrulefill}&\multicolumn{2}{c@{}}{\hrulefill}\\
 & \textbf{Type I error} & \textbf{Sample size} $\bolds{(m, n)}$ & \textbf{Type I error} & \textbf{Sample size} $\bolds{(m, n)}$\\
 & $\bolds{(m, n)=}$& \textbf{needed to reach}& $\bolds{(m, n)=}$ & \textbf{needed to reach} \\
 \textbf{Scenario} &\textbf{(50, 50)}& \textbf{power 0.9} &\textbf{(50, 50)}& \textbf{power 0.9}\\
 \hline
1&0.046&(51, 51) or (92, 46)& 0.050 &(140, 140) or (202, 101)\\
2&0.051&(51, 51) or (92, 46)& 0.049 &(140, 140) or (202, 101)\\
3&0.048&(59, 59) or (100, 50)& 0.050 &(177, 177) or (240, 120)\\
4&0.053&(50, 50) or (92, 46)& 0.052 &(140, 140) or (200, 100)\\
\hline
\end{tabular*}
\end{table}

%f4 ###
\begin{figure}
\tabcolsep=0 pt
\begin{tabular*}{\textwidth}{@{\extracolsep{\fill}}cc@{}}

\includegraphics{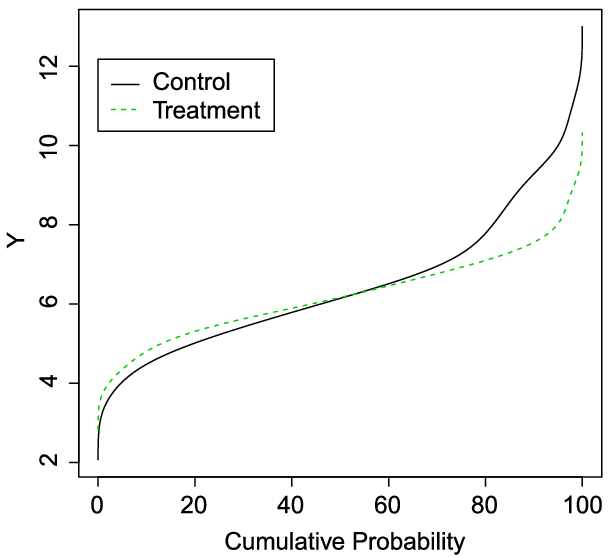}
&\includegraphics{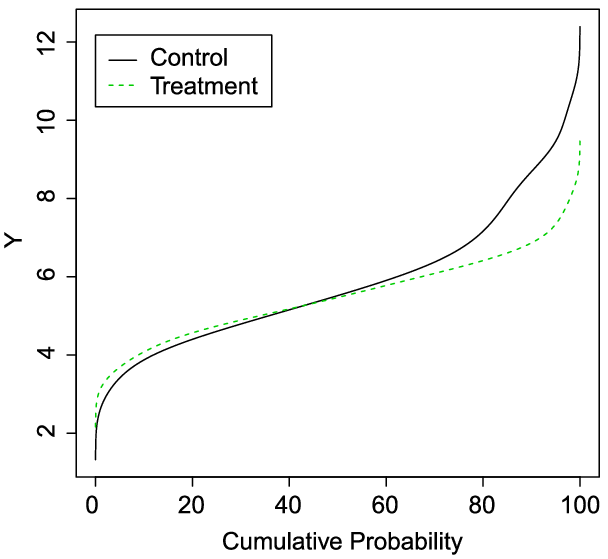}\\
(a)& (b)\\[7 pt]
\multicolumn{2}{@{}c@{}}{
\includegraphics{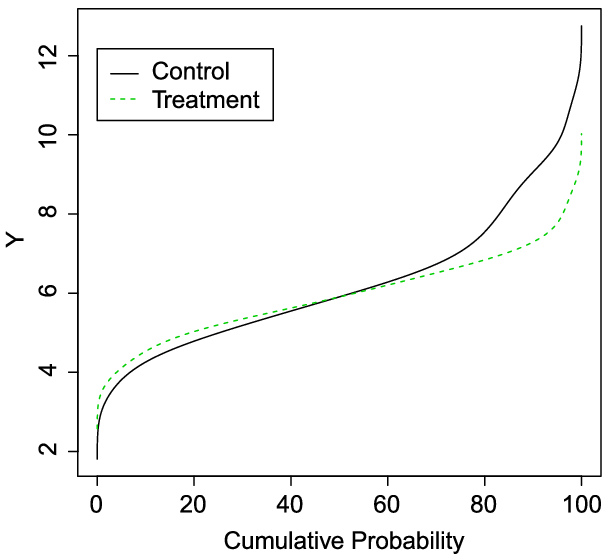}
}\\
\multicolumn{2}{@{}c@{}}{(c)}
\end{tabular*}
\caption{Quantile function plots of the covariate-adjusted outcomes; the adjustments are
made based on regression quantile at (\textup{a}) $\tau=0.5$, (\textup{b}) $\tau=0.75$,
(\textup{c}) $\tau=0.9.$ The diagnostic plots are insensitive
 to the initial choice of $\tau$.}\label{diagnostic}
\end{figure}

The results clearly show the efficiency of the $\operatorname{COVES}$ test. In Scenarios 2--4,
the adjustment of the covariate is important, because the ES test considered in Section~\ref{4.1} would not be valid, and thus it is not presented in this subsection.

%s5 ###
\section{A diagnostic tool for $\operatorname{COVES}$}\label{sec5}

When preliminary or full data are available, it is often helpful to have a simple diagnostic tool
that points to a case in favor of the $\operatorname{COVES}$ test. We suggest examining the quantile function plot, as used in
Figure~\ref{fig1}, but applied to the covariate-adjusted outcomes defined in Section \ref{sec3}. When the quantiles
of covariate-adjusted outcomes from different treatment groups differ mostly in one tail, we
have a clear case in favor of the $\operatorname{COVES}$ test or a similar test that focuses on the tail. In fact, the plot
can also suggest an appropriate level of $\tau$ to be used for $\operatorname{COVES}$. To illustrate this point, we
simulated one data set of size $m=n=60$ from Scenario 3 in Section \ref{4.2} with $\eta=1.35$ in model (\ref{normalmodel}).
Unsure about a good choice of $\tau$, we considered using the covariate-adjusted outcomes from three quantile levels 0.5, 0.75, and 0.9, and
examined the resulting quantile plots in Figure \ref{diagnostic}. No matter which quantile level we started with,
the quantile plots of the covariate-adjusted outcomes look similar, and they all suggest that the $\operatorname{COVES}$ test with $\tau$ around 0.75
would be a good choice. On the other hand, if the quantile functions of different treatment groups show a vertical shift, we would then
favor the $t$-test to the $\operatorname{COVES}$ test.

%s6 ###
\section{Conclusions}\label{sec6}
The proposed $\operatorname{COVES}$ test aims to detect treatment effects that are reflected mostly in the
upper (or lower) tail of the outcome distributions. The test is powered up by the use of
the expected shortfall as
a natural differentiating quantity in such applications.  We find that the regression quantile
methodology is appropriate and convenient for computing the covariate-adjusted expected shortfall
in the test. Our study on the change of the Total Sharp Scores due to different treatments on
rheumatoid arthritis shows that a substantial sample size reduction over the conventional
\textit{t}-test based on linear models can be achieved.

In this paper, we used $\tau=0.75$ in the proposed $\operatorname{COVES}$ test, because it serves
two purposes in the application. First, earlier studies have shown conventional rheumatoid arthritis treatments
are effective for nearly 75\% of the patient population, so it is less meaningful
to detect differences below the 75th percentile. Second, a more effective treatment
should work well for a substantial portion of the patients, so if we set $\tau$ to be
too high in the $\operatorname{COVES}$ test, a significant difference in the upper tail might be difficult
to detect statistically.  Finally, we note that the development of the $\operatorname{COVES}$ test in
this paper was made in response to
the randomized clinical studies on rheumatoid arthritis treatments, but the basic
idea and the methodology clearly generalize to other problems (where tail differences
of possibly other $\tau$ values are) of interest. In general, we suggest using quantile function plots
on covariate-adjusted outcomes as a simple diagnostic tool for suggesting a good
choice of $\tau$.

\begin{appendix}\label{append}

\section*{Appendix: Sketch of proof}

\setcounter{equation}{0}

The following lemma follows directly from the consistency and the Bahadur representation of regression
quantile estimators; see Koenker [(\citeyear{Koenker2005}), Section 4.3] and \citet{HeandShao1996}.

\begin{lemma}\label{lem1}
$\!$If $\{ (Z_i, D_i,  C_i) \}$ is a random sample satisfying (\ref{quant}), $\lim_{m,n\rightarrow\infty}(m+n)^{-1}U_f$ exists,  $E|C_i|^3<\infty$, and $f_i$
are uniformly bounded away from 0 and infinity, then
we have the Bahadur representation on $\hat\gamma$
\[
\hat\gamma-\gamma(\tau)=-U_f^{-1}\sum_i C^*_{i} I(e_{i}<0) +o_p\bigl((m+~n)^{-1/2}\bigr),
\]
and the representation on $\bar e_\tau (d)$
\[
\bar e_\tau (d)- \biggl\{\sum_{D_i=d} I(e_{i}>0)\biggr\}^{-1}\sum_{D_i=d} e_{i}I(e_{i}>0) =o_p\bigl((m+~n)^{-1/2}\bigr),
\]
where $U_f=\sum_i (f_iC_i^{*2}),$ $f_i$ is the conditional density function of $e_i$ given $(D_i, C_i)$ evaluated at 0, and $C_i^* = C_i - N_d^{-1}\sum_{i} C_i I(D_i=d).$
\end{lemma}
%%%%%
%%%%%

\begin{pf*}{Proof of Theorem \ref{theorem}}
By replacing ${\hat e}_i$ in $T_\tau (m,n)$ by $e_i$ and using the results in Lemma \ref{lem1}, we approximate
$T_\tau (m,n)$ by
%%%
%%%
\begin{eqnarray*}
T_\tau^*(m, n)&=&\delta(\tau)+\biggl[\{(1-\tau)m\}^{-1}\sum_{D_i=1} I(e_{i}> 0) e_{i}\\
&&\hphantom{\delta(\tau)+\biggl[}{}-\bigl(\bar C_\tau(1)-\bar C_\tau(0)\bigr)U_f^{-1} \sum_{D_i=1} C_{i}^*I(e_{i}\geq 0)\biggr]\\
&&{}-\biggl[\{(1-\tau)n\}^{-1}\sum_{D_i=0} I(e_{i}> 0) e_{i} \\
&&\hphantom{{}-\biggl[}{}+\bigl(\bar C_\tau(1)-\bar C_\tau(0)\bigr)  U_f^{-1}\sum_{D_i=0} C_{i}^*I(e_{i}\geq 0)\biggr].
\end{eqnarray*}
%%%
%%%
It is clear that $E(T_\tau^*(m, n))=\delta(\tau)=0$ under $H_0$,  and $T_\tau^*(m, n)$
is asymptotically normal, with
%%%
%%%
\begin{eqnarray*}
&&\mbox{$\operatorname{var}$}(T_\tau^*(m, n))\\
&&\qquad=\{(1-\tau)m\}^{-2}\sum_{D_i=1}\bigl(E\{e_i^{2}I(e_i>0)\}-[E\{e_iI(e_i>0)\}]^2\bigr)\\
&&\qquad\quad{}+\tau(1-\tau)\bigl(\bar C_\tau(1)-\bar C_\tau (0)\bigr)^2 U_f^{-2}\sum_i (C_{i}^*)^2\\
&&\qquad\quad{}+\{(1-\tau)n\}^{-2}\sum_{D_i=0}\bigl(E\{e_i^{2}I(e_i>0)\}-[E\{e_iI(e_i>0)\}]^2\bigr).
\end{eqnarray*}
%%%
%%%
Again, by Lemma \ref{lem1} and $T_\tau(m,n)-
T^*_\tau(m,n)=o_p((m+n)^{-1/2})$, the asymptotic
normality of Theorem \ref{theorem} follows.
\end{pf*}

\end{appendix}

\printaddresses

\end{document}